\documentclass[aps,prb,twocolumn,groupedaddress]{revtex4}
\usepackage{graphicx} 

\begin{document}
\title{Hopping Conductivity Beyond The Percolation Regime Probed By Shot-Noise Measurements}
\author{F. E. Camino}
\author{V. V. Kuznetsov}
\author{E. E. Mendez}
\affiliation{Department of Physics and Astronomy, State University of New York at Stony Brook, Stony Brook, New York 11794-3800}
\author{M. E. Gershenson}
\affiliation{Serin Physics Laboratory, Rutgers University, Piscataway, New Jersey 08854-8019}
\author{D. Reuter}
\author{P. Schafmeister}
\author{A. D. Wieck}
\affiliation{Lehrstuhl f\"{u}r Angewandte Festk\"{o}rperphysik, Ruhr-Universit\"{a}t Bochum, Universit\"{a}tsstrasse 150 D-44780 Bochum, Germany}

\date{\today}
\begin{abstract}
We have observed suppression of shot noise in the variable-range hopping regime for a two-dimensional electron gas whose localization length, $\xi$, is controlled by a gate voltage. We have found that the suppression factor $F$ (Fano factor) is approximately inversely proportional to the length of the system $L$ ($F=L_0^*/L$) for a broad range of values of $\xi/L_0^*$, where $L_0^*$ is the characteristic length. In the case $\xi/L_0^*<<1$, we have identified $L_0^*$ with the distance $L_0$ between the hard hops along the sample length. On the other hand, when $\xi$ is of the order of $L_0^*$, we have observed that $L_0^*$ does not agree with $L_0$ calculated from the percolation model of hopping. We attribute this discrepancy to a breakdown of that model and to a reconstruction of the hopping paths.
\end{abstract}


\maketitle

Shot noise in the hopping regime has attracted attention only recently,\cite{Kor,Kuz,Sverdlov} in spite of the fact that shot-noise measurements can shed light on the microscopic processes of electronic transport.\cite{Bla} It has been shown experimentally that the study of shot noise provides important information on electron hopping. \cite{Kuz} In particular, it was found that the noise power spectral density, S, of a two-dimensional (2D) hole gas in a SiGe quantum well is smaller than the Poissonian value of $2eI$, with the Fano factor ($F \equiv  S/2eI$) being inversely proportional to the length of the sample.\cite{Kuz} The proportionality constant was interpreted as a characteristic scale for inhomogeneity of the hopping process.  An analogous reduction of the shot noise was observed recently in GaAs MESFET devices in both two- and one-dimensional (1D) configurations.\cite{Ros}

In the hopping regime, the electrons are localized within the \emph{localization length} $\xi$.  According to the percolation model of hopping,\cite{ES} there is another characteristic length, $L_0 >> \xi$, which represents the distance between the so-called hard hops.  In other words, $L_0$ is the size of cells of the infinite cluster of less-resistive hops, the backbone of the percolative network which carries the current.\cite{ES} At a scale much greater than $L_0$, a sample can be considered homogeneous.

It has been found experimentally that when the percolation model is applicable (when $L >> L_0 >> \xi$) the Fano factor is approximately inversely proportional to the number of hard hops along the sample: $F=L_0/L$, where $L$ is the sample length.\cite{Kuz} This observation is consistent with the theoretical prediction for noise suppression in a 1D hopping system: the Fano factor is $F = 1/N$ for hopping through $N$ identical barriers.\cite{Kor} In order to apply this result to a 2D network, one should treat $N$ as the number of hard hops along the sample length. 

A recent numerical calculation\cite{Sverdlov} has found that in a 2D system $F\propto 1/L^\alpha$ with $\alpha=$ 0.85. Although within the limited data available the experimental result in Ref.~2 is also consistent with an $\alpha=$ 0.85 exponent, that calculation is based on simplifications (e.g., $T=$ 0 K, no e-e interactions) that in some cases may restrict its applicability. Previous experiments\cite{Kuz,Ros} have been analyzed within the framework of the percolation model\cite{ES} using the $F=L_0/L$ dependence, which offers a more intuitive picture and fits the data at least equally well. In the case of a 2D hole system, \cite{Kuz} $L_0$ was determined to be $\sim$1~$\mu$m for the 2-5 $\mu$m-long samples, while $\xi$ was estimated to be $\sim$0.01~$\mu$m.  A similar value for $\xi$ was obtained for shorter (0.2~$\mu$m~$\leq$ L$\leq$~0.4~$\mu$m) GaAs devices.\cite{Ros} In the latter case, however, a deviation from the dependence $F= L_0/L$ was found: the Fano factor saturated below the value of 1, expected when $L_0 \sim L$, for both 2D and 1D configurations.  This result was tentatively attributed to the Coulomb interaction between hopping electrons.\cite{Ros}

Neither of these studies has considered the situation when $L_0$ is of the order of $\xi$.  Under this condition, it is questionable to treat the hopping conductivity as the charge percolation through a 2D network of lumped resistors, which represent individual hops.  It is also unclear what to expect for the Fano factor in such a case. 

\begin{figure}
\includegraphics[width=86mm]{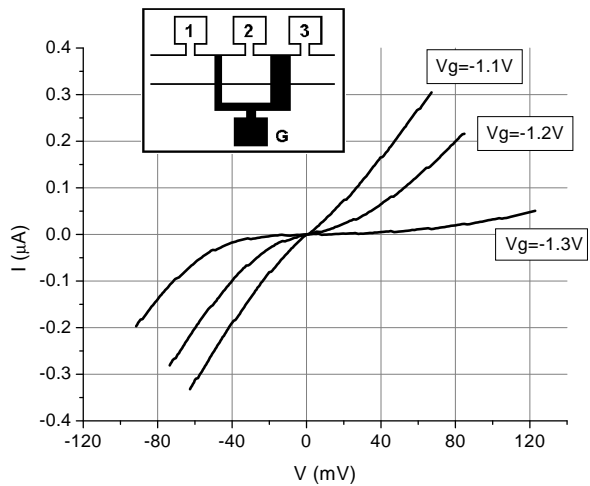}
\caption{\label{IV_5um} Two-terminal current-voltage curves for the section of the sample with a 5~$\mu$m-long gate (contacts 1 and 2 in inset). A relatively large $V_{sd}$ modifies the effective gate voltage, and results in an asymmetry of the I-V curves. Similar asymmetry was observed in the noise measurements (Fig.~\ref{SI_5um}). The 10~$\mu$m gate is selected by using contacts 2 and 3 (see inset).}
\end{figure}

To address this question, we have studied the shot noise of hopping 2D electrons over a wide range of $\xi/L_0$ values, from very small up to values of the order of unity, using a low-mobility 2D electron gas in a Si-doped GaAs structure. A single $\delta$-doped layer of Si donors (concentration 1.3$\times$10$^{12}$ cm$^{-2}$) was separated from the metal-film gate electrode by a 0.1~$\mu$m-thick MBE-grown layer of undoped GaAs . The width of the conducting channel between the source and drain was 100~$\mu$m. Two metal gates across the channel, 5~$\mu$m and 10~$\mu$m long, allowed to deplete the 2D gas and to drive it into the hopping regime (see the inset in Fig.~\ref{IV_5um}).
 
Figure~\ref{IV_5um} shows the source-to-drain current-voltage characteristics for the section of the channel intercepted by the 5~$\mu$m-long gate, at various values of the gate voltage, $V_g=$ -1.1~to~-1.3 V. (Similar curves were obtained for the section intercepted by the 10~$\mu$m-long gate.)  The observed increase of the source-drain resistance with decreasing gate voltage reflects an increasing depletion of carriers under the gate.  The current asymmetry for the two polarities of the source-drain voltage, $V_{sd}$, is a consequence of the electronic configuration employed, in which large positive (negative) values of $V_{sd}$ decrease (increase) the effective gate voltage.

\begin{figure}
\includegraphics[width=86mm]{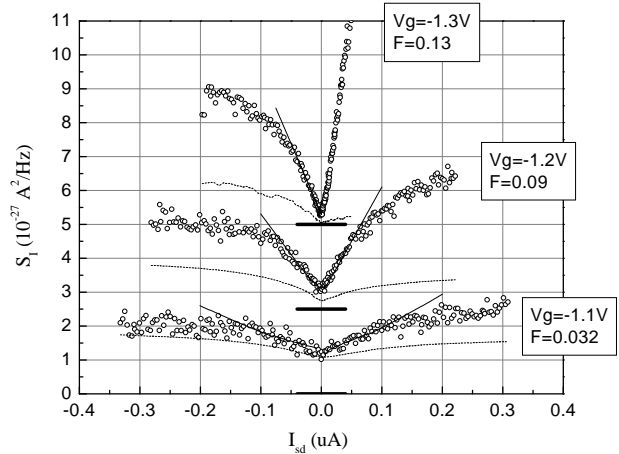}
\caption{\label{SI_5um} Shot noise for the 5~$\mu$m-long gate section of the device shown in the inset in Fig.~\ref{IV_5um}. For clarity, the zero current noise levels (short and thick horizontal lines) for $V_g=$ -1.2~V and $V_g=$ -1.3 V have been shifted by 2.5$\times$10$^{-27}$ A${^2}$/Hz and 5$\times$10$^{-27}$ A${^2}$/Hz, respectively.  The Fano factor, $F$, is an average of the values obtained from the fit of Eq. \ref{eq:SI} to each current polarity, except for $V_g=$ -1.3 V, where only the negative current branch has been considered (to compare it with its equivalent for 10 micron gate length). The dashed curves represent the calculated thermal noise using 4$kTG(I)$, where $G(I)$ is the measured conductance as a function of source-drain current, $I$, for each gate voltage.\cite{ThermalNoise}}
\end{figure}

\begin{figure}
\includegraphics[width=86mm]{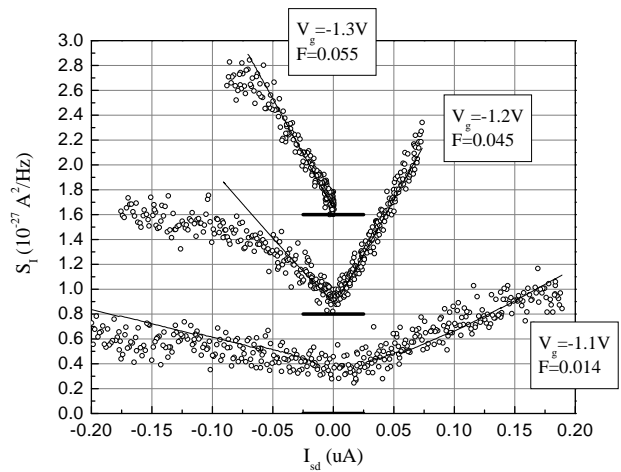}
\caption{\label{SI_10um} Shot noise for the sample with a 10~$\mu$m-long gate, at several gate voltages. The curves for $V_g=$~-1.2 V and $V_g=-$1.3~V have been shifted by 0.8$\times$10$^{-27}$ A${^2}$/Hz and 1.6$\times$10$^{-27}$ A${^2}$/Hz, respectively (short and thick horizontal lines). The positive bias current for $V_g=$~-1.3~V was too small to be resolved with sufficient accuracy. The Fano factor is obtained as described in the caption of Fig.~\ref{SI_5um}.}
\end{figure}

\begin{table*}
\caption{\label{Tabl} Sample parameters for several gate voltages. The values shown in the second and third columns represent the Fano factors for the 5 $\mu$m- and 10 $\mu$m-long sections of the sample, respectively. They were obtained by averaging the Fano factor for negative and positive bias current of the shot noise vs current graphs (see Figs.~\ref{SI_5um} and \ref{SI_10um}), except for $V_g=$~-1.3 V, where only the negative bias branch was considered. The values of $T_0$ are extracted from the fit shown in Fig.~\ref{GvsT}. The localization length $\xi$ was calculated from Eq. \ref{eq:xi}, using $C=$ 1, $\kappa=$ 12.4 for the dielectric constant of GaAs, and the experimental values of $T_0$. The values of $L_0^*$ were calculated from the relation $F=L_0^*/L$ and averaged for the 5~$\mu$m- and 10~$\mu$m-long gates. Finally, the values of $L_0$ were calculated from Eq. \ref{eq:L0}.}
\begin{ruledtabular}
\begin{tabular}{ccccccc}
$V_g$ & $F$ & $F$ & $T_0$ & $\xi$ &  $L_0^*$ & $L_0$ \\
(V) & ($L=$5$\mu$m) & ($L=$10$\mu$m) & (K) & ($\mu$m)  & ($\mu$m) & ($\mu$m) \\ \hline
-1.3 & 0.130$\pm$0.010 & 0.055$\pm$0.005 & 137$\pm$31 & 0.010$\pm$0.002  & 0.60$\pm$0.04 & 0.54$\pm$0.20 \\
-1.2 & 0.090$\pm$0.010 & 0.045$\pm$0.010 & 59.4$\pm$7.8 & 0.023$\pm$0.003  & 0.45$\pm$0.04 & 0.48$\pm$0.10 \\
-1.1 & 0.032$\pm$0.005 & 0.014$\pm$0.004 & 17.0$\pm$0.6 & 0.080$\pm$0.003  & 0.15$\pm$0.02 & 0.40$\pm$0.02 \\
\end{tabular}
\end{ruledtabular}
\end{table*}

The electronic noise of the source-drain current was measured at a frequency of 80 kHz, using a set-up described in Ref.~2 and with the sample immersed in liquid He. The current dependence of the noise power spectral density for several gate voltages is summarized in Figs.~\ref{SI_5um} and \ref{SI_10um}, for the 5~$\mu$m- and 10~$\mu$m-long gates, respectively.  The Fano factor has been extracted from the fit to the expression\cite{Ros} 
\begin{equation}
S_I=F2eIcoth(FeV/2kT).  \label{eq:SI}
\end{equation}
In the $eV> > 2kT/F$ limit, this equation reverts to the familiar $S_I=F2eI$. In our case, however, it is important to use the full expression, especially for $V_g=$ -1.1~V, where $eV\approx 2kT/F$. In Figs. 2 and 3, open circles are experimental points, and solid lines are fits to Eq. \ref{eq:SI}. From an analysis of these data, we draw the following conclusions: (a) In all cases, the shot noise is partially suppressed ($F<1$). (b) For fixed gate length, $F$ increases as $V_g$ becomes more negative, as shown in columns 2 and 3 of Table I for the 5 micron and 10 micron long gates, respectively. This can be interpreted as a decrease of the number of hard hops (increase of $L_0$) as the sample becomes more resistive. (c) The suppression is approximately inversely proportional to the gate length. The values of the proportionality constant, $L_0^*$, are indicated in column 6 of Table I. We note that although the parameters are different, these conclusions are similar to those reached previously using very different materials systems.\cite{Kuz,Ros} This similarity lends additional support to the generality of the underlying physical mechanism and to the simple expression we have used connecting Fano factor and gate length.

\begin{figure}
\includegraphics[width=86mm]{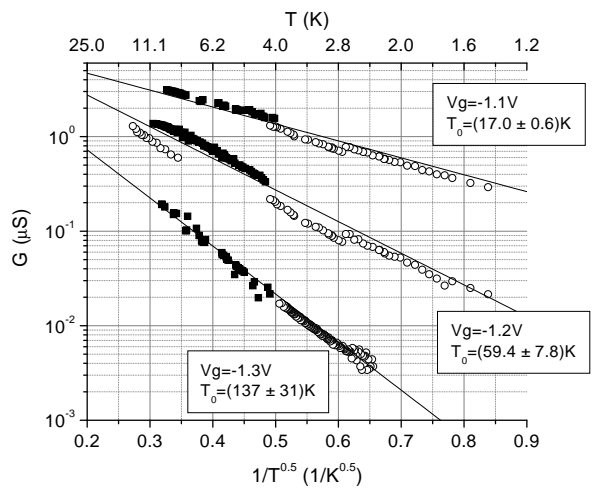}
\caption{\label{GvsT} Temperature dependence of the zero-bias conductance for the sample with a 10~$\mu$m-long gate. Open circles are for measurements in a liquid helium (LH) cryostat that can be pumped down to 1.4 K. Filled squares are data taken with the sample in a LH  transport dewar. Lines are fits to the equation $G(T)\propto exp \; [-(T_0/T)^{0.5}]$ (see Ref. 7). The shift between both sets of data,  especially clear at $V_g=$ -1.2 V, is probably due to differences in the sample resistance after several warm-up and cool-down cycles.}
\end{figure}

In order to compare the experimental results with the theoretical predictions based on the percolation model of hopping,\cite{Kor} we need to estimate the characteristic lengths for the samples studied. In the strongly-localized regime, the temperature dependence of the conductivity (see Fig. \ref{GvsT} and Ref. 7) resembles the Efros-Shklovskii (ES) law for variable-range hopping,\cite{ES} $G(T)\propto exp \; [-(T_0/T)^{0.5}]$. In this case, the localization length can be estimated as
\begin{equation}
\xi=Ce^2/\kappa k_BT_0,    \label{eq:xi}
\end{equation}
where $C$ is a constant of the order of unity,\cite{ES,VK} and $\kappa$ is the dielectric constant. The values of the fitting parameter $T_0$, and $\xi$ calculated from Eq.~\ref{eq:xi}, are shown in Table \ref{Tabl}. Tuning the gate voltage over the range $V_g=$~-1.1~to~-1.3 V results in large variations of $\xi$, from 0.084 $\mu$m to 0.010 $\mu$m. It is worth mentioning that application of the ES model for the hopping with a large localization length ($V_g=$ -1.1 V) is not fully justified (see, e.g., Ref. 9), and we need to treat this estimate with caution.\cite{Ref_Caution}
 
We are now in a position to compare two characteristic lengths, $\xi$ and $L_0^*$. For $V_g=$~-1.3 V, the ratio $\xi/L_0^*$ is much smaller than 1, and the percolation model of hopping holds. We assume that in this case $L_0^*$ coincides with the distance between hard hops, $L_0$ [Ref. 11]:
\begin{equation}
L_0=\xi(T_0/T)^{(\nu+1)/2},  \label{eq:L0}
\end{equation}
where $\nu$ is the critical index of the correlation radius ($\sim$1.3 in 2D). In Table \ref{Tabl} we list the values of $L_0$ obtained from Eq.~\ref{eq:L0} using the tabulated values of $T_0$ and $\xi$. Indeed, for $V_g=$ -1.3 V and -1.2 V, where $\xi/L_0^*< < $ 1, the values of $L_0^*$ and $L_0$ are in good agreement.
 
The situation is different, however, for $V_g=$ -1.1 V, where $\xi/L_0^*=$ 0.53. In this case, $L_0^*$ is a factor of three smaller than the estimate for $L_0$ from Eq. \ref{eq:L0} (see Table \ref{Tabl}). This discrepancy suggests that the percolation model is not applicable when $\xi$ becomes of the order of $L_0^*$. The relation $F=L_0^*/L$ is apparently still valid in this limit, but for a correct interpretation of $L_0^*$ it is necessary to go beyond the percolation model. This implies to take into account not only the backbone of the infinite percolation cluster, but also new hopping trajectories which come into play when $\xi$ becomes comparable to $L_0^*$.

Given the limited (2) number of gate lengths used in our experiment, it is fair to ask whether one could have analyzed the experimental results using instead a $F=(L_0^*/L)^{0.85}$ dependence, as suggested by recent numerical calculations.\cite{Sverdlov}  Such an analysis yields values of $L_0^*$ that are somewhat different from those shown in Table I; however, the large difference between $L_0^*$ and $L_0$ for $V_g=$ -1.1 V remains, indicating that our conclusion is not qualitatively affected by small deviations of the exponent from unity.

In principle, percolation theory could still explain our results if, in analogy with the 3D case, we would assume that the dielectric constant, $k$, increases as the sample becomes more conductive.\cite{SPC} An increase of $k$ by a factor of three at $V_g=$ -1.1 V would give a good match between the predicted $L_0$ from percolation theory and our measured $L_0^*$. This interpretation is appealing since at that gate voltage the sample still seems to follow the ES law (as seen in Fig. \ref{GvsT}). However, it is an open question whether in the 2D system considered here the dielectric constant might grow with increasing localization length. We hope that the experimental results presented here will stimulate the development of a theoretical model applicable for an arbitrary ratio of $\xi/L_0^*$.

In short, we have found that the Fano factor $F$ is approximately inversely proportional to the length of the system $L$ ($F=L_0^*/L$), in a broad range of values of the $\xi/L_0^*$ ratio, where $\xi$ and $L_0^*$ are the localization and characteristic lengths, respectively. In the limit $\xi/L_0^*<<$ 1, we have identified $L_0^*$ with the distance $L_0$ between hard hops along the sample length. On the other hand, when $\xi \sim L_0^*$, we have observed that $L_0^*$ does not agree with $L_0$ calculated from the percolation model of hopping, thus signaling a reconstruction of the hopping paths.

This work has been sponsored by the National Science Foundation Grants Nos. DMR-9804023 (E.E.M.) and DMR-0077825 (M.G.).


\begin{thebibliography}{}
\bibitem{Ref_Caution} We note that, because of the way they were obtained, the values of $\xi$ and $L_0$ in Table I should be considered just as estimates.  Therefore it is more appropriate to focus on the order of magnitude of the $\xi/L_0$ ratio than on its exact value.
\bibitem{ThermalNoise} For $V_g=$ -1.3 V and -1.2 V, the thermal noise component is much smaller than the measured noise, while  for $V_g$= -1.1 V they are of the same order, with the thermal noise smaller, especially for positive currents. In this case, the negative-current branch is closer to the thermal curve because the effective gate voltage is less negative in that region, hence driving the sample into a more metallic state. The calculated Fano factor is largely weighed by the positive current branch, where the measured noise is clearly larger than the thermal component. Same considerations apply to the 10 $\mu$m-long section of the sample.
\end{thebibliography}
\end{document}